\documentclass[%
 aip,
 amsmath,amssymb,
 reprint,
]{revtex4-1}

\usepackage{graphicx}
\usepackage{physics}
\usepackage{siunitx}
\usepackage[utf8]{inputenc}
\usepackage[T1]{fontenc}
\usepackage{mathptmx}
\usepackage{etoolbox}
\usepackage[colorlinks=true,linkcolor=blue, citecolor=blue, urlcolor=blue]{hyperref}

\makeatletter
\def\@email#1#2{%
 \endgroup
 \patchcmd{\titleblock@produce}
  {\frontmatter@RRAPformat}
  {\frontmatter@RRAPformat{\produce@RRAP{*#1\href{mailto:#2}{#2}}}\frontmatter@RRAPformat}
  {}{}
}%
\makeatother
\begin{document}

\preprint{AIP/123-QED}

\title[Nonlinear Nanoresonators for Bell State Generation]{Nonlinear Nanoresonators for Bell State Generation}

\author{Maximilian~A.~Weissflog}
\thanks{These authors contributed equally to this work}
\affiliation{Institute of Applied Physics, Abbe Center of Photonics, Friedrich Schiller University Jena, Albert-Einstein-Straße 15, 07745 Jena, Germany}
\affiliation{Max Planck School of Photonics, Hans-Knöll-Straße 1, 07745 Jena, Germany}

\author{Romain~Dezert}
\thanks{These authors contributed equally to this work}
\affiliation{Laboratoire Matériaux et Phénomènes Quantiques, Université Paris-Cité, CNRS, 10 Rue Alice Domon et Léonie Duquet, F-75013 Paris, France}

\author{Vincent~Vinel}
\affiliation{Laboratoire Matériaux et Phénomènes Quantiques, Université Paris-Cité, CNRS, 10 Rue Alice Domon et Léonie Duquet, F-75013 Paris, France}

\author{Carlo~Gigli}
\affiliation{Laboratoire Matériaux et Phénomènes Quantiques, Université Paris-Cité, CNRS, 10 Rue Alice Domon et Léonie Duquet, F-75013 Paris, France}

\author{Giuseppe~Leo}
\affiliation{Laboratoire Matériaux et Phénomènes Quantiques, Université Paris-Cité, CNRS, 10 Rue Alice Domon et Léonie Duquet, F-75013 Paris, France}

\author{Thomas~Pertsch}
\affiliation{Institute of Applied Physics, Abbe Center of Photonics, Friedrich Schiller University Jena, Albert-Einstein-Straße 15, 07745 Jena, Germany}
\affiliation{Fraunhofer Institute for Applied Optics and Precision Engineering IOF, Albert-Einstein-Straße 7, 07745 Jena, Germany}

\author{Frank~Setzpfandt}
\affiliation{Institute of Applied Physics, Abbe Center of Photonics, Friedrich Schiller University Jena, Albert-Einstein-Straße 15, 07745 Jena, Germany}
\affiliation{Fraunhofer Institute for Applied Optics and Precision Engineering IOF, Albert-Einstein-Straße 7, 07745 Jena, Germany}

\author{Adrien~Borne}
\email{adrien.borne@univ-paris-diderot.fr}
\affiliation{Laboratoire Matériaux et Phénomènes Quantiques, Université Paris-Cité, CNRS, 10 Rue Alice Domon et Léonie Duquet, F-75013 Paris, France}

\author{Sina~Saravi}
\email{sina.saravi@uni-jena.de}
\affiliation{Institute of Applied Physics, Abbe Center of Photonics, Friedrich Schiller University Jena, Albert-Einstein-Straße 15, 07745 Jena, Germany}

\begin{abstract}
Entangled photon states are a fundamental resource for optical quantum technologies and investigating the fundamental predictions of quantum mechanics. Up to now such states are mainly generated in macroscopic nonlinear optical systems with elaborately tailored optical properties. In this theoretical work, we extend the understanding on the generation of entangled photonic states towards the nanoscale regime, by investigating the fundamental properties of photon-pair-generation in sub-wavelength nonlinear nanoresonators. Taking materials with Zinc-Blende structure as example, we reveal that such systems can naturally generate various polarization-entangled Bell states over a very broad range of wavelengths and emission directions, with little to no engineering needed. Interestingly, we uncover different regimes of operation, where polarization-entangled photons can be generated with dependence on or complete independence from the pumping wavelength and polarization, and the modal content of the nanoresonator. Our work also shows the potential of nonlinear nanoresonators as miniaturized sources of biphoton states with highly complex and tunable properties.
\end{abstract}

\maketitle

\section{Introduction}

Entangled photonic states, and among them entangled photon-pair states, are one of the main resources for realizing optical quantum technologies, from quantum communication and computation \cite{Flamini19} to quantum imaging.\cite{Basset19} Nonlinear optical systems have been used as the dominant approach for generation of entangled photon-pair states.\cite{Wang21} 
In general, when using parametric nonlinear optical processes, it is more natural to generate photon pairs that are entangled in the spectral degree of freedom, while this is not necessarily the case in the spatial or polarization degrees of freedom. 
Bulk nonlinear crystals can generate polarization-entangled photon pairs, yet with very limited control over their spatial and spectral properties.\cite{Kwiat95} More elaborate schemes like using two back-to-back crystals\cite{Kwiat99} or placing a crystal in a Sagnac loop\cite{Hentschel09} were used to add versatility to such bulk systems. Nanostructured and integrated nonlinear systems on the other hand can increase the possibilities for generation of entangled states by engineering the optical modes or the nonlinearity profile of the system. This has been implemented for example by simultaneously satisfying two phase-matching processes in a ridge waveguide\cite{Horn13} or a biperiodically poled waveguide\cite{Herrmann13} for generating polarization-entangled states, or using a photonic crystal waveguide\cite{Saravi17} or coupled waveguides\cite{Solntsev14} for generating path-entangled states.
Careful post-processing after generation using well-aligned polarization beamsplitters and birefringent waveplates is another way of generating polarization-entangled photon-pair states.\cite{Martin10}
All such schemes require very careful engineering of the optical structure and/or the system's alignment to create indistinguishability between two quantum states, such that their superposition can create a maximally entangled Bell state with two photons.

Recently, there have been several experimental demonstrations of photon-pair generation\cite{sharapovaNonlinearDielectricNanoresonators2023} and manipulation\cite{wangQuantumMetasurfaceMultiphoton2018,solntsevMetasurfacesQuantumPhotonics2021} in much smaller nonlinear systems, with at least one subwavelength-sized dimension, like thin-films \cite{okoth2019microscale,okoth2020idealized,santiago2021entangled, sultanov2022flat}, nanowires \cite{saerensBackgroundFreeNearInfraredBiphoton2023}, dielectric nanoresonators \cite{Marino19}, and metasurfaces.\cite{Santiago21,santiago2022,zhang2022spatially,son2023photon} For instance Ref.\cite{sultanov2022flat} demonstrated spectrally broadband generation of tunable polarization-entangled photon-pairs from a sub-wavelength nonlinear film, not restricted by phase-matching conditions. These experiments suggest that nanoscale nonlinear systems are a promising platform for the generation of photon pairs with a wide range of control over their various degrees of freedom, different than what can be achieved in bulk or even waveguide or high Q-factor resonator systems.

\begin{figure}[t!]
\centering
\includegraphics[width=1\linewidth]{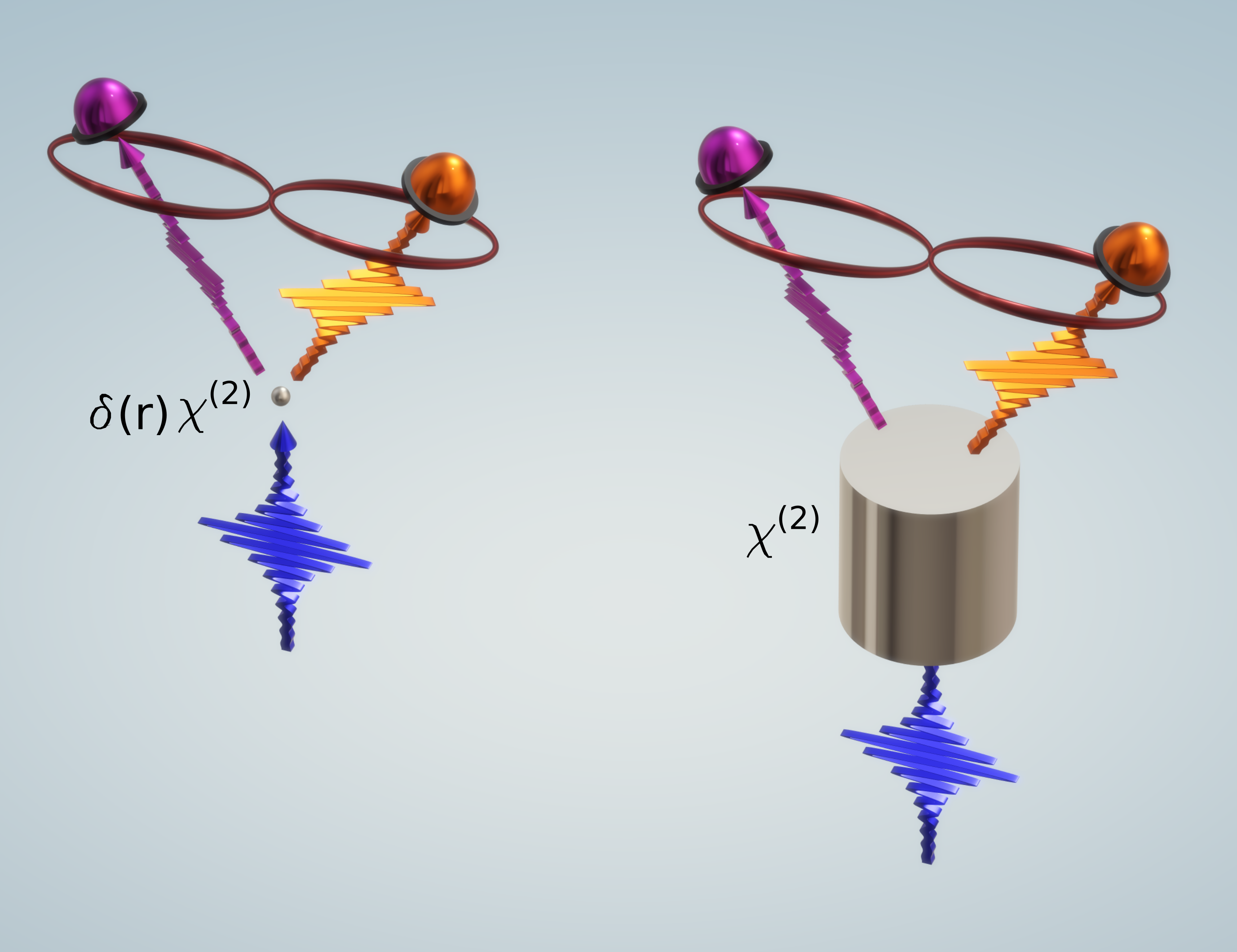}
\caption{Schematic of photon pair-generation in a point-like nonlinear system (left) and a cylindrical nonlinear dielectric nanoresonator (right), where by irradiating the nonlinear structure with a pump beam (blue), entangled signal and idler photons (orange and purple) can be generated and detected in the far-field.}
\label{fig:Fig_1}
\end{figure} 

With these motivations, in this work, we investigate the mechanism of photon-pair generation in nanoscale nonlinear systems, to extend the understanding about entangled photon-pair generation to the nanoscale regime.
We analyze the process of spontaneous parametric down-conversion (SPDC), in which a material with second-order nonlinearity is pumped by an optical beam of angular frequency $\omega_p$. This probabilistically generates pairs of signal and idler photons, respectively, with angular frequencies $\omega_s$ and $\omega_i$ that fulfill the energy conservation condition $\omega_p=\omega_s+\omega_i$.
To this end, we develop a new numerical approach, which allows us to uncover new regimes of entanglement generation in nanoscale systems.
Taking materials with ZincBlende structure as example, we show that such nonlinear systems can naturally generate maximally polarization-entangled states over very broad frequencies and emission directions without special modal engineering. We demonstrate this for point-like nonlinear nanoparticles as well as for nanoresonators (see Fig.~\ref{fig:Fig_1}).
In particular, we highlight a regime of operation, where by using a certain class of detection configurations, maximally polarization-entangled photon pairs can be detected in all emission directions and over a wide range of wavelengths. This effect is fully independent of the pumping wavelength, polarization, and the modal content of the nonlinear nanoscale system.
We call this a "protected" regime of entanglement generation, as generation of maximally polarization-entangled states show a strong robustness to the variation of many of the system parameters.
In contrast to this protected regime, we also show that in a different operation regime, strongly influenced by the detection configuration, the entanglement properties can be fully controlled through the pumping properties and modal interferences in the nonlinear nanoscale system.

For our theoretical demonstrations, we first investigate the case of a point-like (infinitely small) nonlinear structure, see left side of Fig.~\ref{fig:Fig_1}. Then we study a more experimentally relevant system, namely dielectric nanoresonators with Mie-type resonances, shown on the right side of Fig.~\ref{fig:Fig_1}.
Through a combination of the Green's function quantization method and the quasinormal-mode expansion, we create a method that allows us to numerically investigate the rich physics of pair-generation in such nanoresonators. Throughout this work, we focus on the large class of nonlinear materials with a cubic lattice and point group $\mathrm{\Bar{4}3m}$, a crystal shape in the literature and throughout this manuscript referred to as "Zinc-Blende" structure. Examples are III-V semiconductors like GaAs, GaP, and InP, and II-VI semiconductors like ZnTe, ZnSe, and ZnS. They have second-order nonlinear susceptibility tensors of the form $\chi^{(2)}_{\alpha \neq \beta \neq \gamma}$, where the indices $\{ \alpha,\beta,\gamma \}$ refer to the directions along the axes of the nonlinear crystal $\{ x_c,y_c,z_c \}$. Due to the cross-polarized nature of their $\chi^{(2)}$ tensor, such nonlinear materials are naturally suited for generation of polarization-entangled Bell states.

\section{Results}\label{sec2}

\subsection{Bell state generation by a point-like nonlinear source}

\begin{figure*}[t!]
\centering\includegraphics[width=1\linewidth]{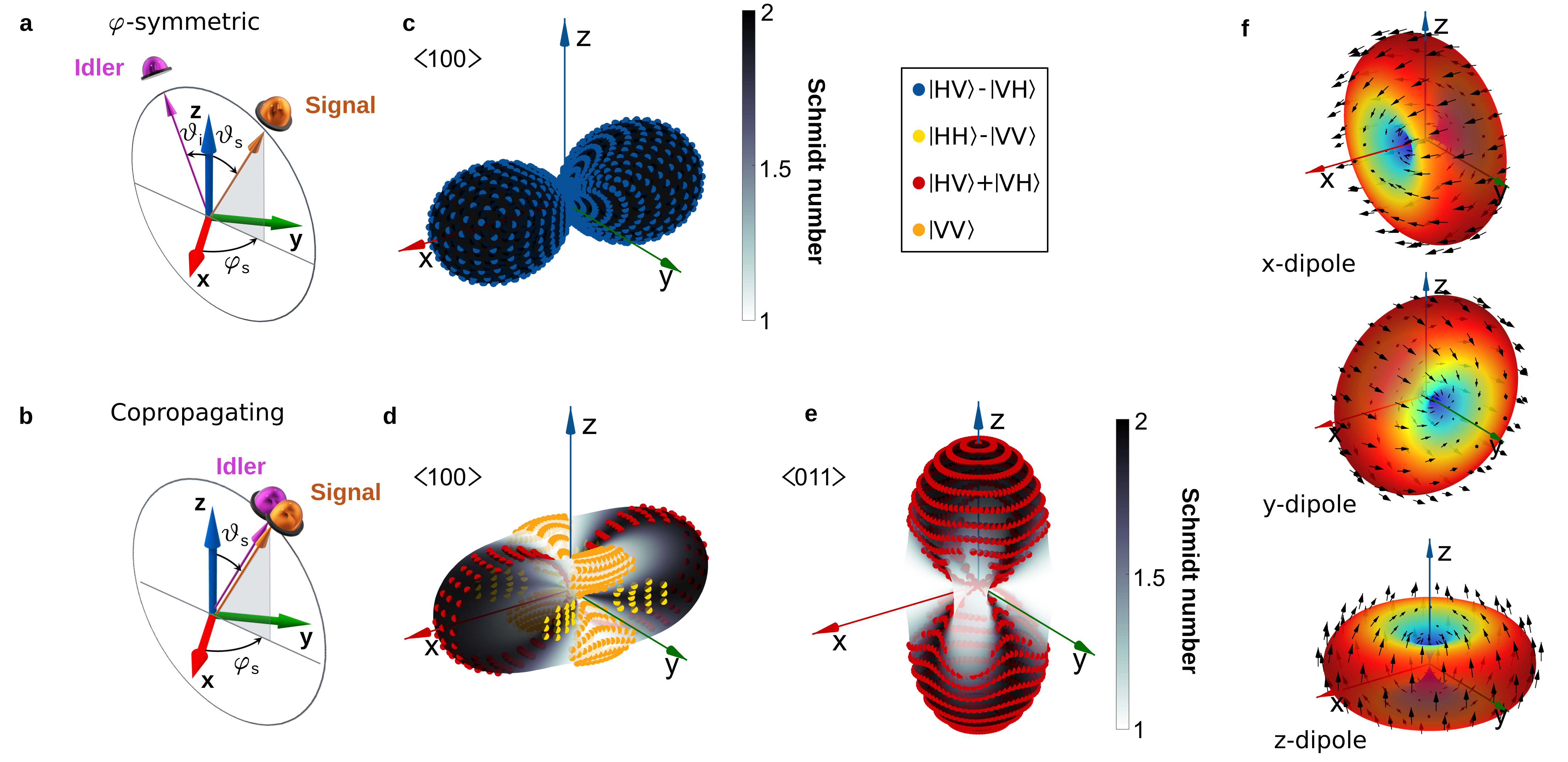}
\caption{\textbf{Properties of the biphoton states generated in a Zinc-Blende point-like nonlinear source.} \textbf{a-b} Visualization of the different detection configurations. (\textbf{a}) $\varphi$-symmetric ($\theta_i=\theta_s$ and $\varphi_i=\varphi_s+\pi$) and (\textbf{b}) copropagating ($\theta_s=\theta_i$ and $\varphi_s=\varphi_i$). \textbf{c-e} Mapping of the photon-pair state properties in all detection angles, where each case corresponds to the detector configuration shown on the left side of the same line (\textbf{a})-(\textbf{b}). The directional pattern indicates the coincidence detection rate ($d^4N_{pair}/dt d\omega_s d\Omega_s d\Omega_i $), the color map is the Schmidt number and the coloured dots mark several quantum states identified with a fidelity $F>0.9$.
(\textbf{c}) and (\textbf{d}) ore obtained for a ${<}100{>}$-oriented Zinc-Blende crystal and (\textbf{e}) for a ${<}011{>}$ orientation.
\textbf{f} Far-field radiation pattern of $x$-, $y$- and $z$-oriented electric dipoles, respectively. The far-field polarization vectors $\Vec{P}_{x}(\theta,\varphi)$, $\Vec{P}_{y}(\theta,\varphi)$ and $\Vec{P}_{z}(\theta,\varphi)$ of the respective $x$-, $y$- and $z$-electric dipoles are marked with black arrows.
}
\label{fig:Fig_2}
\end{figure*}

We start by considering a nanoparticle in the Rayleigh scattering regime with a radius much smaller than the wavelengths involved in the SPDC process. We model such a system in the lowest order approximation by a point-like nonlinear susceptibility, $\chi^{(2)}_{\alpha \beta \gamma} \delta(\vec{r})$, which does not disturb the linear properties of the background medium.
For our calculations, we use a Green's function (GF) method that can treat photon-pair generation in arbitrary dispersive and open optical systems\cite{poddubnyGenerationPhotonPlasmonQuantum2016} (for more details see Appendix~\ref{sec:GF_method_photon_pairs}).
We first consider a Zinc-Blende crystal with orientation ${<}100{>}$ as the nonlinear material for the point-like source. The crystal axes $\{ x_c,y_c,z_c \}$ are aligned with the lab coordinates $\{ x,y,z \}$. The pump beam is monochromatic, polarized along the $x$-direction and propagating along -$z$.
Throughout this work, we consider a detection scheme where the generated signal and idler photons are selectively registered in the far-field of the source, as shown in Fig.~\ref{fig:Fig_2}(a)-(b), by placing small-area detectors in a particular combination of directions $\{\theta_s,\varphi_s\}$ for the signal and $\{\theta_i,\varphi_i\}$ for the idler photon. Additionally, bandpass spectral filters are placed in each angular channel, centered around the desired signal and idler wavelengths, $\lambda_s$ and $\lambda_i$, respectively.
The quantum state of the generated photon pairs is mapped in the far-field by moving over all the possible generation angles for the signal photon. Each corresponding idler photon is detected at a fixed angular relation with respect to the signal. We initially focus on a particular configuration, which we refer to as the $\varphi$-symmetric detection configuration, where for every signal photon at $\{\theta_s,\varphi_s\}$, the idler photon is detected at $\theta_i=\theta_s$ and $\varphi_i=\varphi_s+\pi$, as shown schematically in Fig.~\ref{fig:Fig_2}(a). For each possible direction of the signal photon (which correspondingly fixes the idler detection direction), we calculate both the coincidence detection rate and the polarization state of the biphoton quantum state.
Notice that for a photon-pair state, by fixing the direction and frequency degrees of freedom, polarization remains as the only degree of freedom for the quantum state. While the former degrees of freedom are used to distinguish the signal and idler photons, the corresponding density matrix for the polarization quantum states is retrieved using polarization tomography. For more details see Appendix~\ref{sec:Q_state_tomography} and supplementary section S1.
For each photon-pair state, we also quantify the degree of polarization entanglement by calculating the Schmidt number (see Appendix~\ref{sec:Q_state_tomography}), where a value of 1 corresponds to a fully un-entangled polarization state and a value of 2 corresponds to a maximally entangled state.

The result for the $\varphi$-symmetric case is shown in Fig.~\ref{fig:Fig_2}(c) as a function of the signal propagation angles $\{\theta_s,\varphi_s\}$. The directional pattern indicates the coincidence detection rate, which is maximum along the $x$-axis and zero in the $yz$-plane. See supplementary section S2 for an example derivation of the coincidence detection rate for a point-like nonlinear source. The color of the pattern indicates the Schmidt number for each photon-pair state in the detection direction of the signal photon. We also mark the found polarization state at many discrete points. In our analysis, the $H$ and $V$ polarizations for each photon are set as the local polarization basis that is orthogonal to that photon's propagation direction. This choice of orthogonal polarization basis is further motivated and described in appendix~\ref{sec:Q_state_tomography} and supplementary section S1. Note that for a point-like source, the biphoton emission pattern and polarization state in the far-field will be completely independent of the frequencies of the signal and idler photons, and there is only a weak frequency-dependence on the total generation rate.

Interestingly, as can be seen in Fig.~\ref{fig:Fig_2}(c), the exact same maximally polarization-entangled state of $\ket{\psi}=\ket{H}_s\ket{V}_i-\ket{V}_s\ket{H}_i=\ket{HV}-\ket{VH}$ is found for all propagation directions. In other words, in the $\varphi$-symmetric detection configuration, a ${<}100{>}$-oriented Zinc-Blende point-like SPDC source generates the same polarization-entangled Bell state over the whole space. 
Notice that the Schmidt number is 2 in all directions, emphasizing the generation of a maximally entangled state.
To understand this effect, we point out that with an $x$-polarized pump and point-like ${<}100{>}$-oriented nonlinear source, the nonlinear tensor elements $\chi^{(2)}_{zyx}$ and $\chi^{(2)}_{yzx}$ participate in the SPDC process. 
Therefore, two cross-polarized electric-dipole sources with $y$- and $z$-polarizations, are responsible for creation of the fields at the signal and idler wavelengths. The specific two-lobed detection pattern in Fig.~\ref{fig:Fig_2}(c) for the rate of coincidences is a result of the multiplication of the intensity of each point on the donut-shaped radiation pattern of the $y$-polarized dipole by its $\varphi$-symmetric corresponding point on the radiation pattern of the $z$-polarized dipole. For a better understanding, we also show the classical far-field radiation patterns and polarizations for $x$-, $y$-, and $z$-polarized electric dipoles in Fig.~\ref{fig:Fig_2}(f).
In fact, two processes can happen, where in process (1) the signal photon propagating along the angle $\{\theta_s,\varphi_s\}$ is generated by a $y$-polarized dipole such that the signal photon has the polarization $\Vec{P}_{y}(\theta_s,\varphi_s)=\Vec{P}_{y,s}$. In this case, the idler photon propagating along the angle $\{\theta_i,\varphi_i\}$ is generated by a $z$-polarized electric dipole and the idler photon has the polarization $\Vec{P}_{z}(\theta_i,\varphi_i)=\Vec{P}_{z,i}$. The polarization vectors $\Vec{P}_{y}(\theta,\varphi)$ for the $y$-dipole and $\Vec{P}_{z}(\theta,\varphi)$ for the $z$-dipole are visualized in Fig.~\ref{fig:Fig_2}(f). In process (2) the situation is reversed, i.e. the role of $y$- and $z$-dipole for signal and idler emission is exchanged. Hence, for a ${<}100{>}$ Zinc-Blende tensor with $x$-polarized pump the two-photon state propagating along a pair of directions will generally have the form $\ket{\psi}=\ket{\Vec{P}_{z,s}}\ket{\Vec{P}_{y,i}}+\ket{\Vec{P}_{y,s}}\ket{\Vec{P}_{z,i}}$. For photons propagating along the $\pm x$-directions, it is straightforward to see that this leads to an entangled state: take the $+x$-direction for the signal photon. Process-(1) generates a $y$-polarized signal photon along the $+x$-direction and a $z$-polarized idler photon along the $-x$-direction and process-(2) generates a $z$-polarized signal photon along the $+x$-direction and a $y$-polarized idler photon long the $-x$-direction. Hence, we get exactly orthogonal polarizations between the signal photons as well as orthogonal polarizations between the idler photons in the two superimposed processes.
For other detection directions this is actually not the case. In other words, for a general angle $\{\theta_s,\varphi_s\}$, the signal from process-(1), generated by a $y$-polarized dipole, does not have an orthogonal polarization to the signal photon from process-(2), generated by a $z$-polarized dipole. Yet interestingly, the sum of the two processes still yields the $\ket{HV}-\ket{VH}$ state. This is a direct consequence of the interference of the two $\varphi$-symmetric processes, which completely cancels out their individual contribution to the un-entangled part of the biphoton state, such that only the maximally polarization-entangled part of the biphoton state remains. In order to show this, we decompose each of the signal and idler photons $\ket{\Vec{P}_{y}}$ (the single-photon state generated by a $y$-polarized dipole) as $\ket{\Vec{P}_{y}} = \alpha\ket{\Vec{P}_{z}}+\beta\ket{\Vec{P}_{z}^\perp}$. Here, $\ket{\Vec{P}_{z}^\perp}$ is a single-photon state in the same propagation direction with a polarization perpendicular to $\ket{\Vec{P}_{z}}$ (the single-photon state generated by a $z$-polarized dipole) such that $\braket{\Vec{P}_{z}}{\Vec{P}_{z}^\perp}=0$.
The $\alpha$-coefficient for this vector projection is simply derived as $\alpha(\theta,\varphi)= \braket{\Vec{P}_{z}}{\Vec{P}_{y}} = \Vec{P}_{z}^*(\theta,\varphi)\cdot\Vec{P}_{y}(\theta,\varphi)$. This indicates the dot-product between the far-field electric-field vectors generated by a $z$-polarized and a $y$-polarized dipole for propagation angles $\{\theta,\varphi\}$. All single-photon states are normalized. In this way, the two-photon state can be expanded into $\ket{\psi}= (\alpha_i+\alpha_s)\ket{\Vec{P}_{z,s}}\ket{\Vec{P}_{z,i}}+ \beta_i\ket{\Vec{P}_{z,s}}\ket{\Vec{P}_{z,i}^\perp}+ \beta_s\ket{\Vec{P}_{z,s}^\perp}\ket{\Vec{P}_{z,i}}$. Note that, while $\Vec{P}_{z,s}^*\cdot\Vec{P}_{z,s}^\perp=\Vec{P}_{z,i}^*\cdot\Vec{P}_{z,i}^\perp=0$ by definition, $\Vec{P}_{z,s}$ and $\Vec{P}_{z,i}^\perp$ as well as $\Vec{P}_{z,i}$ and $\Vec{P}_{z,s}^\perp$ are not necessarily orthogonal. This form of $\ket{\psi}$ allows to easily see that the condition for having a maximally entangled polarization state is to have $\alpha_s+\alpha_i=\alpha(\theta_s,\varphi_s)+\alpha(\theta_i,\varphi_i)=0$ (which also results in $\lvert\beta_i\rvert=\lvert\beta_s\rvert$), and the condition for having a fully un-entangled state is to have $\lvert\alpha_i\rvert=\lvert\alpha_s\rvert=1$.
For this particular case, the $\alpha$-coefficient can be derived to be $\alpha(\theta,\varphi) = \frac{-\mathrm{sign}[\sin(2\theta)] \sin(\varphi)}{\sqrt{1+\tan^2(\theta)\cos^2(\varphi)}}$, compare supplementary section S3 for the full derivation.
It is now easy to see that for the $\varphi$-symmetric case $\alpha_s=\alpha(\theta_s,\varphi_s)=-\alpha_i=-\alpha(\theta_i=\theta_s,\varphi_i=\varphi_s+\pi)$ is satisfied for all possibilities of $\{\theta_s,\varphi_s\}$.

It is very important to notice that with an ideal point-like source, this analysis is in fact independent of the wavelength of the signal and idler photons, as the derived $\alpha$-coefficient is wavelength-independent. Hence, in this operation regime with $\varphi$-symmetric detection and fixed pump polarization, any pair of signal and idler frequencies that satisfy the conservation of energy will produce the same maximally polarization-entangled state as shown in Fig.~\ref{fig:Fig_2}(c). The result is an extremely broadband, hyper-entangled state, that is spectrally entangled by virtue of energy conservation and at each pair of signal and idler frequencies produces the same polarization-entangled state in all detection directions fulfilling the $\varphi$-symmetry.

Next to this detection configuration, a different operation mode exists for point-like systems that allows to control the extent of entanglement. Take the same scenario of a ${<}100{>}$ crystal and an $x$-polarized pump, but in a copropagating detection configuration where $\theta_s=\theta_i$ and $\varphi_s=\varphi_i$, as shown schematically in Fig.~\ref{fig:Fig_2}(b). In this case, $\alpha_i+\alpha_s=0$ is not satisfied for all directions, hence different degrees of entanglement are obtained along different directions. This can be seen in the 3D map of the quantum state for the copropagating configuration, shown in Fig.~\ref{fig:Fig_2}(d).
Here, the coincidence radiation pattern is slightly different, as now the intensity of each point on the radiation pattern of the $y$-polarized dipole is multiplied by the same point on the radiation pattern of the $z$-polarized dipole. A varying extent of entanglement is found depending on the detection direction. Aside from maximally-entangled pairs on the $xz$- and $xy$-planes (for which $\alpha_i=\alpha_s=0$ is satisfied), fully un-entangled states $\ket{\psi}=\ket{VV}$ with a Schmidt number of 1 are generated on the $yz$-plane, where $\lvert\alpha_i\rvert=\lvert\alpha_s\rvert=1$.

In the configurations considered up to now, the photon-pairs are emitted mainly perpendicular to the pump propagation direction. Additionally, from a practical viewpoint, a scheme with pumping and collection of the pairs along the same axis would be useful. This can for instance be achieved by rotating the crystal to ${<}011{>}$ orientation. In this case, the crystal axis $x_c$ coincides with the lab axis $x$, and the $y_c$- and $z_c$- crystal axes are rotated by 45 degrees with respect to the $y$- and $z$- axes.\cite{xuForwardBackwardSwitching2020,weissflogFarFieldPolarizationEngineering2022} 
With a $y$-polarized pump, the nonlinear tensor elements $\chi^{(2)}_{xyy}$ and $\chi^{(2)}_{yxy}$ participate in the pair-generation. Here again, two processes can happen, where in process (1) the signal photon is generated by the $x$-polarized dipole and the idler photon by the $y$-polarized dipole, while in the process (2) it is vice versa. The combination of the $x$- and $y$-dipoles allow then for emission along the $z$-axis, see Fig.~\ref{fig:Fig_2}(e). For copropagating detection, a maximally polarization-entangled state of $\ket{\psi}=\ket{HV}+\ket{VH}$ is obtained over a broad range of directions in the whole $xz$ and $yz$ planes, for which $\alpha_i=\alpha_s=0$. In this case, the $\alpha$-coefficient is $\alpha(\theta,\varphi)=\Vec{P}_{y}^*(\theta,\varphi)\cdot\Vec{P}_{x}(\theta,\varphi)$ (see supplementary section S3 for its derivation). In addition, emission patterns and quantum state maps identical to those shown in Fig.~\ref{fig:Fig_2}(e) are also obtained considering counter-propagating or $\varphi$-symmetric detection for the ${<}011{>}$ orientation and $y$-polarized pump.

\subsection{Bell state generation by a finite-volume nonlinear nanoresonator}

\subsubsection{QNM-analysis of pair-generation}

The question now is: Would a realistic nanoscale nonlinear structure of finite volume also sustain the fundamental ability of a point-like system to generate polarization-entangled Bell states? 

\begin{figure*}[t]
\centering\includegraphics[width=1\linewidth]{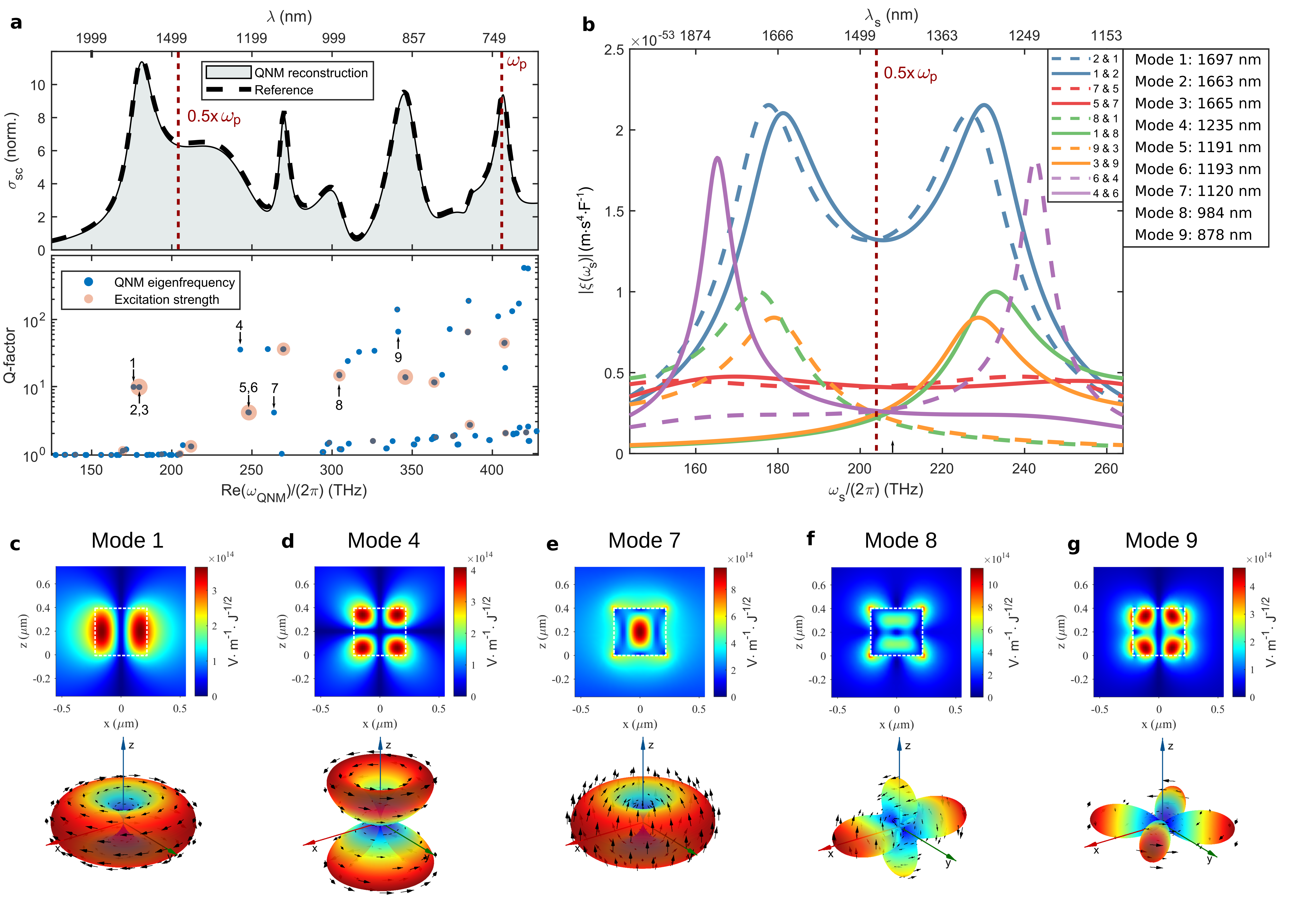}
\caption{\textbf{Modal analysis of linear scattering cross-section and SPDC process.}
\textbf{a}~Upper panel: QNM reconstruction of the linear scattering cross-section of a cylindrical nanoresonator for linearly polarized plane wave excitation incident along the z-axis (grey shaded area). The dashed lines show the scattering cross-section obtained from a FEM full-wave simulation as reference. Lower panel: map of eigenfrequencies used for the reconstruction of the scattering cross-section. Real value and associated mode Q-factor of the complex eigenfrequency are marked with blue dots. Red-dots radii are proportional to the excitation strength of each mode by a plane-wave in the linear scattering problem. \textbf{b} Dispersive modal overlap factor for SPDC excited at $\lambda_p=\SI{735}{nm}$. The contribution of the nine dominant modes is shown. \textbf{c}-\textbf{g} QNM nearfields (upper row) and corresponding far-fields (lower row). For the magnetic-type dipole in panel~(c) and the electric-type dipole in panel~(e) only the $z$-oriented modes are shown.
}
\label{fig:Fig_3}
\end{figure*}

The fundamental difference between the two cases is that the point-like nonlinear system can radiate the signal and idler quantum fields only with an electric-dipole pattern, while a finite-volume system can support a large number of radiative resonant modes.
Each eigenmode can exhibit complex radiation patterns comprising in general a combination of both electric and magnetic multipole radiations, including dipoles, quadrupoles, and higher order spherical harmonics.\cite{wuIntrinsicMultipolarContents2020}
In principle, a large number of modes from a nanoresonator can participate in the pair-generation process, since for the subwavelength-sized system with broad resonances, no strong restrictions from phase- or resonance matching are imposed.
In the following, we show that despite these seemingly complicated dynamics of pair-generation in realistic nonlinear nanoresonators, they can naturally generate maximally entangled photon-pair states. In different operation regimes, the output state is either robust against propagation direction and pump properties or highly tunable.
For this demonstration, we consider one of the simplest and most widely used forms of such systems, a cylinder-shaped dielectric nanoresonator immersed in a homogeneous free-space background, shown schematically in Fig.~\ref{fig:Fig_1}.
Here we choose a height $h=\SI{400}{nm}$ and radius $r=\SI{220}{nm}$ to have the fundamental resonances spectrally located in the near-infrared region.
We use a constant, lossless relative permittivity of $\varepsilon=11.15$ and a quadratic nonlinear susceptibility of $\chi^{(2)}=\SI{200}{pm/V}$ for the resonator This is close to the value of Al$_{0.18}$Ga$_{0.82}$As in the near-infrared range, a widely used material platform for nanostructured nonlinear quantum optical systems.\cite{Marino19,mobini22} Given the generally radiative/leaky nature of a nanoscale system, its eigenmodes have complex-valued eigenfrequencies and are referred to as quasinormal modes (QNMs).\cite{Lal18} For describing the pair-generation dynamics, we resort to the GF method, where this time we expand the GF of the finite-sized nanoresonator into its QNMs basis. Compare Appendix~\ref{sec:GF_QNM_method} for the analytical derivation and Appendix~\ref{sec:computing_near_far_fields} for an explanation of the numerical implementation. Note that the non-dispersive material combined with a homogeneous background is only chosen to not overcomplicate the demonstration model and to keep the focus on the fundamental physics of pair-generation in nanoscale systems. In general, QNM expansions can also be applied to structures made from dispersive metallic\cite{yanRigorousModalAnalysis2018} or dielectric\cite{sauvanQuasinormalModesExpansions2021} materials and with inhomogeneous backgrounds. We point out that pair-generation in dielectric nanoresonators has been investigated in a few previous works,\cite{poddubny2018nonlinear,olekhno2018spontaneous,nikolaeva2021directional} with a focus on the spatial and/or spectral properties, yet without an in-depth investigation of polarization entanglement. In Fig.~\ref{fig:Fig_3}(a) we reconstruct the linear scattering cross-section of the nanocylinder for excitation with a linearly polarized plane wave incident along the $-z$-direction (upper panel) using a set of nearly 400 normalized QNMs. Their associated complex eigenfrequencies are displayed on the lower panel of Fig.~\ref{fig:Fig_3}(a). The excellent agreement between the reconstructed scattering response and the result from a direct full-wave scattering calculation (thick dashed line) shows that the QNM set accurately describes the resonator properties. Throughout the remaining manuscript, we consider a linearly polarized plane-wave excitation along $-z$ with pump intensity of $I_0=1\times10^{9}~\mathrm{W/m^2}$, where the pump field distribution in the nanocylinder, $\Vec{E}_p(\vec{r})$, is directly obtained from a linear full-wave simulation.

An important property of the finite-sized resonator is the frequency dispersion of the resonant modes which directly influence the spectral properties of the generated photon pairs.
We consider a ${<}100{>}$ Zinc-Blende tensor and excitation at one of the strong resonances of the structure, $\lambda_p=\SI{735}{nm}$ (vertical dashed line on Fig.~\ref{fig:Fig_3}(a)). The contribution of each pair of QNMs $m,n$ to the down-conversion process is captured by the dispersive modal overlap factor $\xi_{m,n}(\omega_s, \omega_p)$, see Eq.~\eqref{eq:overlap_xi} in appendix C for the explicit definition. To illustrate this, we evaluate $\xi_{m,n}(\omega_s, \omega_p)$ in Fig.~\ref{fig:Fig_3}(b) for the five pairs of modes with the strongest contribution to the down-conversion process. In Fig.~\ref{fig:Fig_3}(c)-(g), we present the near-field profiles and far-field patterns (calculated at $\omega_p/2$) of these main contributing QNMs. The modes labeled 1 and 7, respectively, exhibit the characteristic features of a z-oriented magnetic and electric dipole. Although their field profiles are not shown here, the QNMs labelled 2, 3, 5 and 6 have similar features as $x$- and $y$-oriented magnetic or electric dipoles, respectively.
The modal overlap factor $\xi_{m,n}(\omega_s, \omega_p)$ for a pair of modes $m,n$ is enhanced when either the signal frequency $\omega_s$ is close to the resonance of one of the QNMs $\omega_s=\widetilde{\omega}_{n}$ or the idler frequency $\omega_i=\omega_p-\omega_s$ is in resonance with a QNM $\omega_i=\widetilde{\omega}_{m}$.

Unless when both resonances are spectrally close to the degenerate SPDC frequency, the resulting spectral contribution of a mode combination will show two peaks, e.g. for the modal combinations 1\&2 in Fig.~\ref{fig:Fig_3}(b).
In case both of the interacting modes have a low Q-factor, see e.g. modes 5, 6 and 7 in Fig.~\ref{fig:Fig_3}(a), the broadband nature of these resonances will lead to an almost flat spectral contribution (e.g. modal combinations 5\&7). For the combination of a high- and low-Q resonance, only one pronounced peak at the higher Q-factor resonance appears. 
Next to the shown pairs of most excited modes, a much larger number of QNMs contribute in the down-conversion process, as expected from a low-Q cavity with broadband resonances. 
In fact, considering an ensemble of \textit{N} QNMs, \textit{N}$^2$ combination of modes have in principle to be evaluated to retrieve the resonator response. In practice however, this number is reduced since QNMs of purely numerical origin, associated to the perfectly matched layers (PMLs) and therefore exhibiting fields mainly localized within the PML,\cite{yanRigorousModalAnalysis2018} do not couple to the pump field and have negligible weight in the SPDC response. 
On the other hand, the tensor configuration might also forbid the excitation of certain families of QNMs. In Fig.~S2 of the supplementary document we show a convergence study which reveals that the photon state properties can already be approximated with excellent accuracy by considering about 50 modes, only an eighth of the calculated QNM dataset. This fast convergence is quite valuable for a fast preliminary screening of resonator properties.

\subsubsection{Entangled two-photon far-fields from a <100> AlGaAs Nanoresonator}

\begin{figure*}[t!]
\centering\includegraphics[width=1\linewidth]{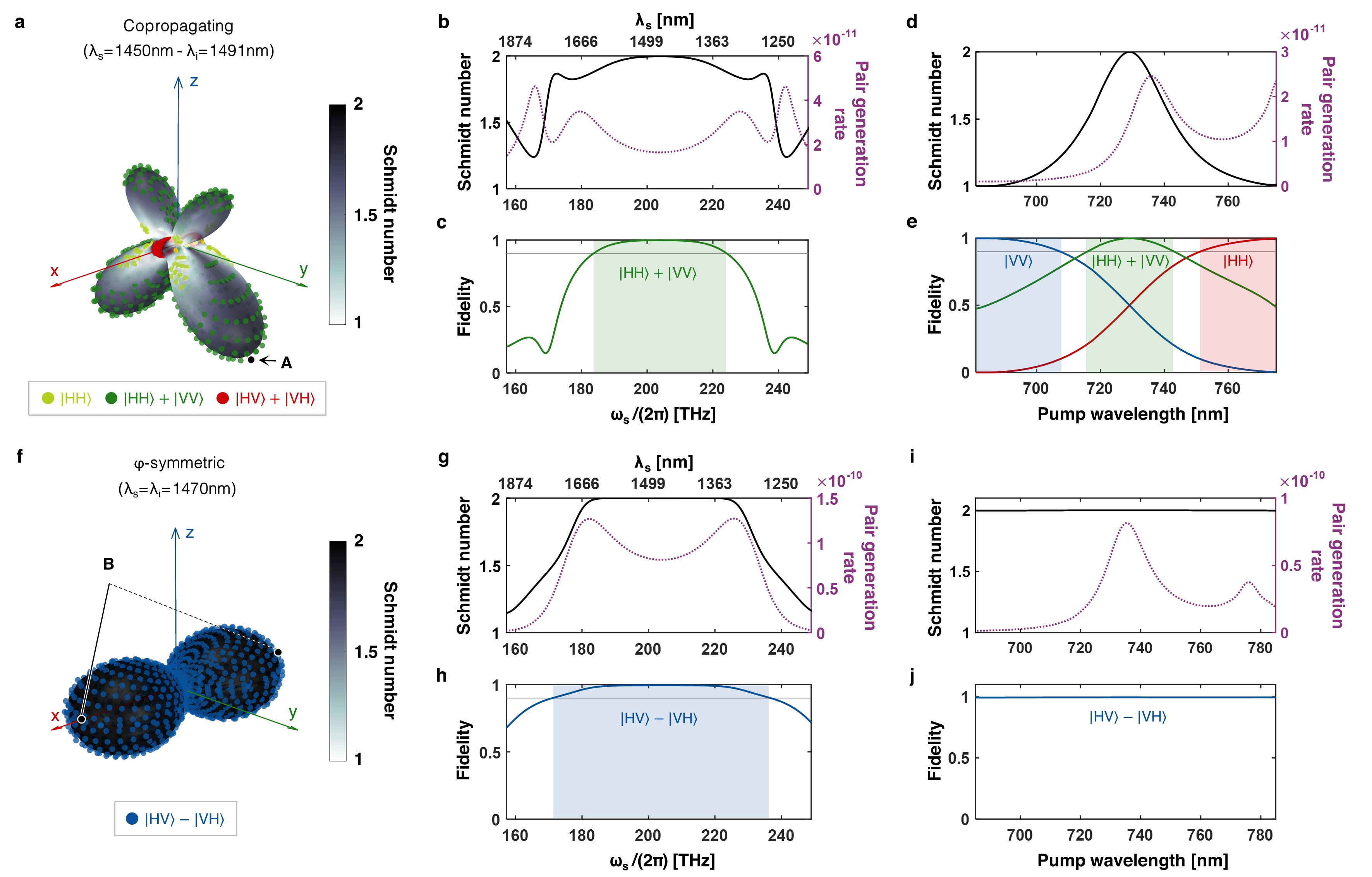}
\caption{\textbf{Properties of the biphoton states generated in a ${<}100{>}$ AlGaAs nanocylinder.}
\textbf{a(f)} Mapping of the photon-pair state properties in all detection angles, for a pump wavelength $\lambda_p=735$ nm in (\textbf{a}) copropagating detection configuration with non-degenerate photon pairs ($\lambda_s=1450$ nm, $\lambda_i=1491$ nm) and (\textbf{f}) $\varphi$-symmetric detection configuration with degenerate photon pairs ($\lambda_s=\lambda_i=1470$ nm). 
The color scale indicates the Schmidt number of the calculated states for each direction of the pairs. The colored dots identify states having a fidelity above 0.9 with the mentioned separable or maximally-entangled states.
\textbf{b-e} and \textbf{g-j}
Evolution of the Schmidt number, the generation rate at a specific emission angle ($d^4N_{\mathrm{pair}} / dt d\omega_s d\Omega_s d\Omega_i$), and the fidelity to specific polarization states, as a function of the generated signal wavelength and the pump wavelength.
The (\textbf{b-e}) panels correspond to the biphoton state in the copropagating configuration at a specific emission angle ($\theta =135^{\circ}$, $\varphi=90^{\circ}$, see point A on panel \textbf{a}). Panels (\textbf{b,c}) are considering a fixed pump wavelength of $\lambda_p=730$ nm.
Panels (\textbf{d,e}) are calculated for non-degenerate photon pairs with a signal wavelength $\lambda_s=2\lambda_p-20$ nm (the idler wavelength is fixed according to the energy conservation).
Panels (\textbf{g-j}) correspond to the biphoton state in the $\varphi$-symmetric configuration with the signal photon propagating along a specific direction 
($\theta_s=90^{\circ},\varphi_s=0^{\circ}$, see point B on panel \textbf{f}, where the corresponding idler emission angle along $\theta_s=90^{\circ},\varphi_s=180^{\circ}$ is also marked).
Panels (\textbf{g, h}) are considering a fixed pump wavelength of $\lambda_p=735$ nm.
Panels (\textbf{i, j}) are calculated for degenerate photon pairs. The colored rectangles in (\textbf{c}), (\textbf{e}), and (\textbf{h}) identify the spectral bands for which the fidelity to the mentioned Bell state is larger than $0.9$ (threshold represented by the grey horizontal line).}
\label{fig:Fig_4}
\end{figure*}

Panels (a) and (f) of Fig.~\ref{fig:Fig_4} present the spatial photon-pair emission diagrams of the nonlinear nanocylinder pumped by an $x$-polarized plane-wave at a wavelength of $\lambda_p=\SI{735}{nm}$. Two detection schemes are shown: copropagating (a) and $\varphi$-symmetric detection (f). For the latter case (f), degenerate photon pairs are considered and signal and idler modes are distinguished by the spatially different detector positions. For the copropagating configuration (a), signal and idler are spectrally distinguished by considering photons approximately $\SI{20}{nm}$ from degeneracy ($\lambda_s=\SI{1450}{nm}$ and $\lambda_i=\SI{1490.6}{nm}$).
For both cases, the calculated Schmidt numbers reach values of 2 and different Bell states are obtained with a high fidelity ($F > 0.9$). The striking difference is, however, that also for the finite-sized resonator, changing the detection configuration switches between either a regime with tunable state generation (a) or emission of the same "protected" polarization state over all directions (f).

Starting with analyzing the former, the spatial shape of the emission pattern in (a) differs significantly from that of Fig.~\ref{fig:Fig_2}(d) obtained for the point-like source under the same detection configuration, since the finite-size resonator now exhibits a richer modal response. The generation pattern of the resonator is determined by the spatially varying interference between all QNMs excited at the signal and idler wavelengths.
However, unchanged is the fact that no emission occurs along the $y$- and $z$- axis, originating from the property of the ${<}100{>}$-oriented Zinc-Blende tensor, which only permits coupling of cross-polarized field components.

Furthermore, we see that the degree of entanglement and the generated polarization state are varying depending on the detection direction.
We focus on a particular direction that maximizes the biphoton emission rate in the copropagating case (a), at $\theta=135^\circ,\varphi=90^\circ$ marked with "A" in the figure. For this direction, we report for a fixed $\lambda_p=\SI{730}{nm}$ the evolution of the Schmidt number, the generation rate, and the fidelity to the polarization state $\ket{HH}+\ket{VV}$ as a function of the signal photon frequency (panels (b-c)). We see a spectrally extremely broadband generation of this Bell state, with a fidelity of more than 0.9 over about $\SI{300}{nm}$ of bandwidth for the signal and idler photons. Given that such photon-pair quantum states are naturally entangled in the frequency degree of freedom, such a system creates entanglement in polarization and spectrum, yet in a certain emission direction.
Since the interference between the excited QNMs govern the polarization along a certain direction, the modal content of the nanoresonator can be leveraged to control the generated polarization state, e.g. by tuning the frequency and polarization of the pump beam. In the same configuration as before (with copropagating photons at $\theta=135^\circ,\varphi=90^\circ$, and nondegenerate pairs such that $\lambda_s=2\lambda_{p}-\SI{20}{nm}$), we now follow in Fig.~\ref{fig:Fig_4}(d)-(e) the evolution of the Schmidt number, the generation rate, and the fidelity as the pump wavelength varies. We see that the polarization state generated in this direction switches between two fully separable states and a maximally entangled state. Beyond simply impacting the polarization state, the tuning of the pump properties also affects the efficiency of the photon-pair generation. As can be seen on Fig.~\ref{fig:Fig_4}(b,d), the generation rate can be resonantly enhanced due to the combined excitation of modes near the pump, signal and idler wavelengths.

A remarkable property of this simple nanoresonator is that, analogous to the point-like source, a second operation mode exists. A single "protected" maximally-entangled polarization state is generated when simply changing the detection scheme to a $\varphi$-symmetric configuration. In this case, pairs detected in all possible directions are found in the state $\ket{\psi}=\ket{HV}-\ket{VH}$, as can be seen in Fig.~\ref{fig:Fig_4}(f). Moreover, this maximally entangled state is preserved for a wide spectral band of non-degenerate pairs (over $\SI{400}{nm}$ with a fidelity above 0.9), as can be seen in Fig.~\ref{fig:Fig_4}(g)-(h). Hence, considering this $\varphi$-symmetric configuration, the nanoresonator creates a highly spatially multimode biphoton state entangled in polarization and spectral degrees of freedom.
Such a behavior strongly contrasts with that of a periodic arrangement of interacting nanoresonators, elaborately designed for reaching a Bound State in the Continuum resonance. There, generation of polarization-entangled biphoton states have been predicted over narrow spectral regions and narrow, non-trivial emission angles.\cite{parry2021enhanced} Furthermore it is worth noting that this mode of entanglement generation also differs from that observed for bulk crystals e.g. with type-II noncollinear phase-matching.\cite{Kwiat95} For such sources, polarization entanglement is only observed when symmetrically placing the two detectors at the two intersection points of the emission cones of the signal and idler photons of orthogonal polarization. In contrast, we revealed here that the nonlinear nanoresonator generates maximally entangled states for any $\varphi$-symmetric positioning of the detectors.

Remarkably, the generation of this maximally entangled state over all directions is also completely independent from the pump excitation wavelength, as can be seen in Fig.~\ref{fig:Fig_4}(i)-(j). This means that despite the changing weight of the excited QNMs, the modes always collectively satisfy an interference condition that results in the creation of a Bell state.Equivalently, one could also change the excitation strength of the QNMs by scaling the resonator size while keeping the aspect ratio and pump wavelength constant and would still generate the same, entangled quantum state.
This condition is only perfectly satisfied at the degeneracy wavelength, where the fidelity peaks in Fig.~\ref{fig:Fig_4}(h), and slowly drops away from the degeneracy wavelength.
This can be understood by referring to our analytical treatment for the point-like source with the $\alpha$-coefficients. The $\alpha$-coefficients essentially describe the emission properties of the system from the position of the point-like source to the detection positions for the signal and idler photons. This depends on the far-field GF of the system at each of the signal and idler frequencies. For the simple case of the point-like source in free-space, the far-field GF of the system is the same at the signal and idler frequencies, aside from a phase factor, which for the $\varphi$-symmetric configuration allows to satisfy the condition $\alpha_s+\alpha_i=0$ for all emission directions.

In the case of a nanocylinder, the system GFs at the signal and idler frequencies are only equal at the degeneracy wavelength, but not for non-degenerate wavelengths. 
Although we cannot identify an analytical condition for predicting maximal entanglement in the nanocylinder case, it is nonetheless clear that a similar effect is responsible here. All the separable polarization components in the biphoton state cancel out for every emission direction of the $\varphi$-symmetric configuration.
By virtue of the very broad spectral response of the nanoresonator around our wavelengths of interest for the signal and idler photons, the change in the GF of the system is weak. This results in a slow degradation of maximal entanglement away from the degeneracy wavelength, leading to only a slow decrease of the Schmidt number away from degeneracy in Fig.~\ref{fig:Fig_4}(g).

We also point out that this all-directional entanglement generation in the $\varphi$-symmetric configuration is not only independent of the pump wavelength, but also the pump polarization. In Fig.~S5 of the supplementary document we show that when rotating the linearly polarized pump in the $xy$-plane, the $\varphi$-symmetric spatial photon-pair emission pattern rotates around the $z$-axis together with the direction of the pump polarization. The polarization state remains the same.

\subsubsection{Absolute pair-generation rate}

An important question is the absolute efficiency of the pair-generation process, which is obtained by integrating the differential pair-rate $d^4N_\mathrm{pair}/\left(dt d\Omega_i d\Omega_s d\omega_s\right)$ Eq.~\eqref{N_solid_without2pi} over the entire solid angle and summing over all detected polarization directions.

\begin{figure}[ht!]
\centering\includegraphics[width=1\linewidth]{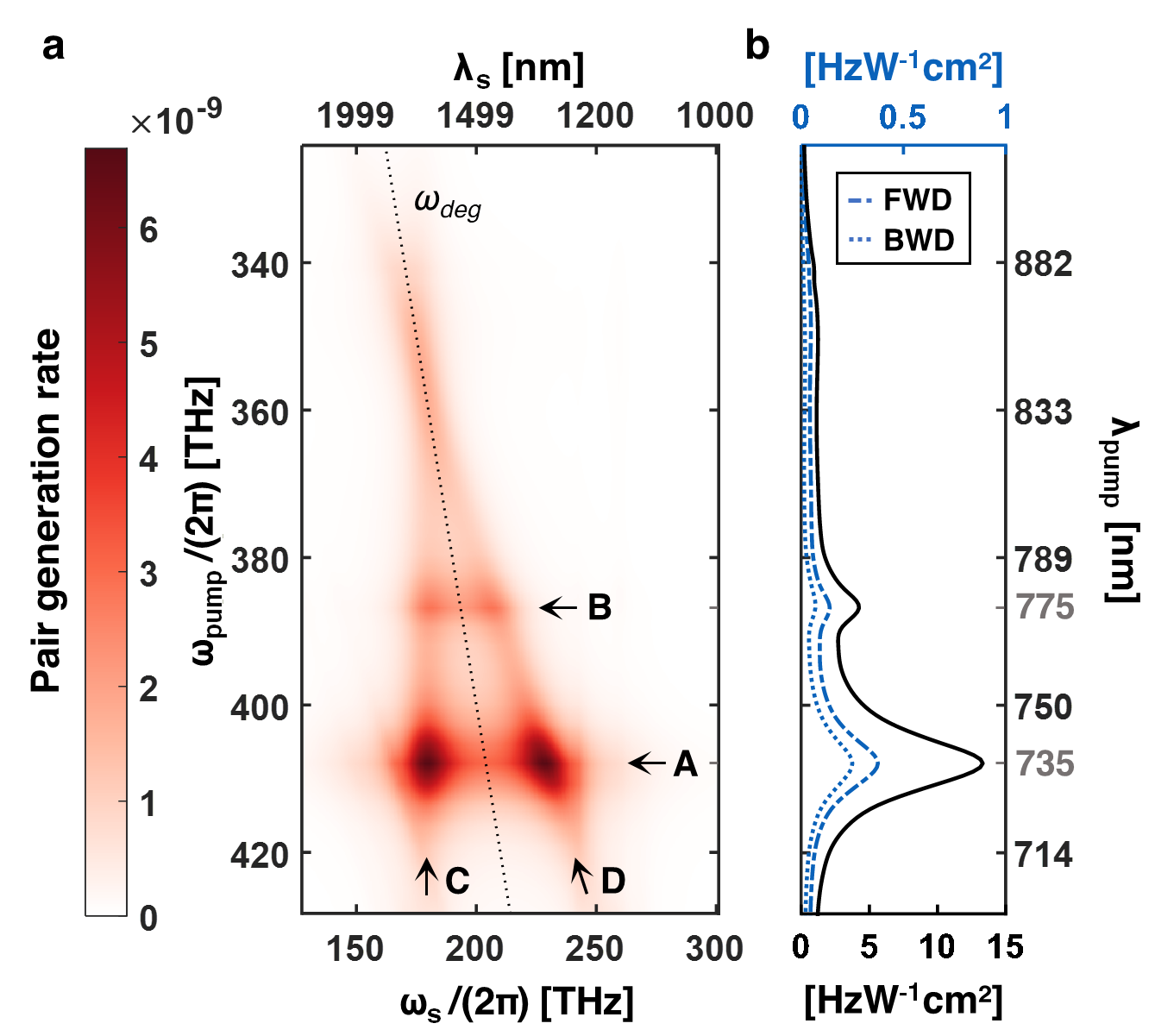}
\caption{\textbf{Pair generation spectra of a ${<}100{>}$ AlGaAs nanocylinder.}
\textbf{a} Evolution of the signal photon spectrum ($d^2N_{pair}/\left( dt d\omega_s \right)$) with the pump wavelength. The spectral degeneracy of the pairs is indicated by the grey dotted line. The A,B,C,D arrows respectively indicate: $\lambda_p=\SI{735}{nm}$, $\lambda_p=\SI{775}{nm}$, $\lambda_s=\SI{1665}{nm}$, $\omega_i=\omega_{deg}-\omega_{s}\vert _{\lambda_s=\SI{1665}{nm}}$. \textbf{b} Evolution of the normalized and spectrally integrated generation rate ($dN_{\mathrm{pair}}/\left( dt I_0 \right)$) with the pump frequency when the pairs are detected in a numerical aperture $NA=0.8$ in the forward direction 
(blue dashed line), in a numerical aperture $NA=0.8$ in the backward direction (blue dotted line), and in the whole space without any spatial filtering (black continuous line - bottom axis). The pump intensity is constant and taken as $I_0=\SI{1e9}{W/m^2}$ in the calculation.
}
\label{fig:Fig_5}
\end{figure}
Panel (a) of Fig.~\ref{fig:Fig_5} shows the signal photon spectrum ($d^2N_{pair}/\left( dt d\omega_s \right)$) as a function of the pump wavelength and demonstrates how the modal content of the resonator both leads to enhancement of the generation rate and influences the bandwidth of the generated photons.
Specifically, a significant increase of the generation rate is provided by narrow resonances at $\lambda_p=\SI{735}{nm}$ and $\SI{775}{nm}$ fundamental wavelengths (horizontal markers A and B). Furthermore, the dipolar-type QNMs 1-3 near $\SI{1665}{nm}$ (Fig.~\ref{fig:Fig_3}(a) and \ref{fig:Fig_3}(c)), also increase the pair-generation rate (vertical marker C). Energy conservation also provides generation enhancement along the correlated spectral line indicated by marker D. The degree of detuning between the degenerate SPDC wavelength and the fundamental QNMs 1-3 of the resonator can be used to control the bandwidth of the generated pairs, where in our example increasing $\omega_p$ significantly broadens the SPDC spectrum. The theoretical maximum pair-rate generated by the ${<}100{>}$ AlGaAs nanocylinder is $\frac{1}{I_0} \frac{dN_{\mathrm{pair}}}{dt}=\SI{13}{Hz\cdot cm^2\cdot W^{-1}}$, obtained by integrating over the SPDC spectrum from $\SI{1}{\mu m}$ to $\SI{2.3}{\mu m}$ at excitation wavelength $\lambda_p=\SI{735}{nm}$ (see Fig.~\ref{fig:Fig_5}(b)). It is very important to note that this value corresponds to an idealized scenario, where pairs emitted over the entire solid angle and a large spectral bandwidth can be detected. In an experiment, the detected rate will likely be lower since it strongly depends on conditions such as spectral bandwidth of the detector, numerical aperture of the objective lens etc. Therefore, we also consider in Fig.~\ref{fig:Fig_5}(b) a common experimental detection scenario, where only coincidences in a finite numerical aperture of $\mathrm{NA}=0.8$ in either only forward (transmission) or only backward (reflection) direction are collected. The spectrally integrated detection rate shown in Fig.~\ref{fig:Fig_5}(b) demonstrates that the qualitative spectral response of the pair-generation in both directions is similar, but is with a maximum value of $\approx\SI{0.4}{Hz\cdot cm^2\cdot W^{-1}}$ approximately 35-times lower than the theoretical maximum. This significant difference is partly because no counter-propagating photon pairs are registered in the collection schemes restricted to only forward or backward collection. However, it also largely stems from the ${<}100{>}$ oriented Zinc-Blende tensor which does not allow efficient pair emission along the z-axis. This shows that even such a nanoscale structure can in principle be a highly efficient source of entangled photon pairs, but the experimental conditions are crucial. The normalized rate of about $\approx\SI{0.4}{Hz\cdot cm^2\cdot W^{-1}}$ for collection with a limited numerical aperture is comparable with rates experimentally measured for resonant nanostructures, when taking into account the respective total detection efficiencies of the experimental setups (including collection efficiency, optical losses, detector efficiency etc.).\cite{santiago2022,zhang2022spatially} Our method can be used to specifically optimize both, nanostructures and experimental schemes, for higher pair-generation and -collection efficiencies.

\subsubsection{Biphoton state tuning by pump control}

\begin{figure*}[t!]
\centering\includegraphics[width=1\linewidth]{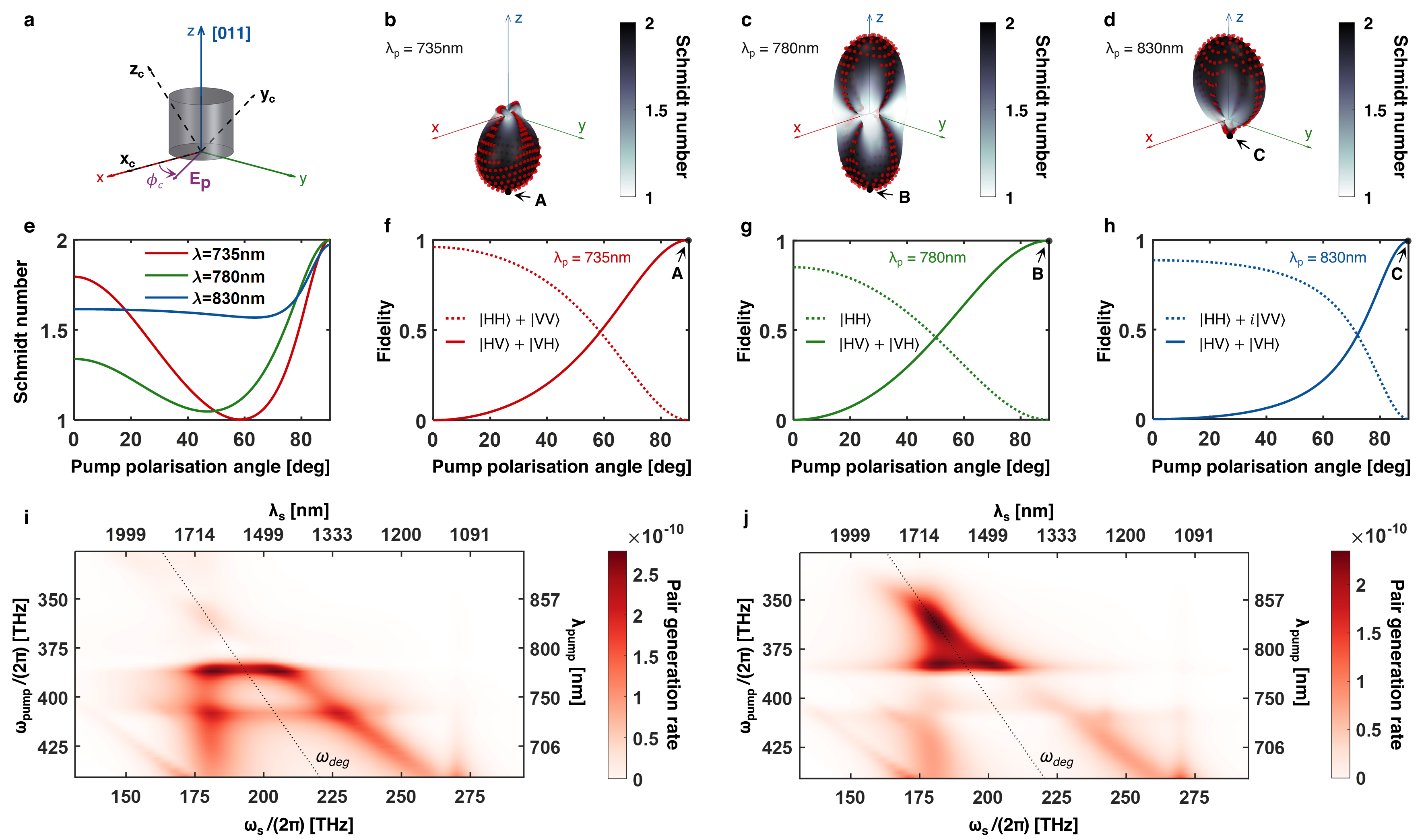}
\caption{\textbf{Biphoton state switching by tuning the pump polarization and wavelength in a ${<}011{>}$ AlGaAs nanocylinder.}
\textbf{a} Schematic view of the crystalline axes for ${<}011{>}$ AlGaAs. The pump plane wave polarization makes an angle $\phi_c$ with the $x=x_c$ axis. 
\textbf{b-d} Mapping of the photon-pair state properties in all detection angles, calculated for copropagating configuration with spectrally non-degenerate pairs ($\lambda_s=2\lambda_p-20$ nm), with a $y$-polarized plane-wave pump ($\phi_c=90^{\circ}$) at (\textbf{b}) $\lambda_p=735$ nm, (\textbf{c}) $\lambda_p=780$ nm, and (\textbf{d}) $\lambda_p=830$ nm. 
The color scale indicates the Schmidt number of the states calculated in each direction and the red dots identify states having a fidelity $F > 0.9$ with the $\ket{HV}+\ket{VH}$ state. 
\textbf{e} Evolution of the Schmidt number calculated for non-degenerate pairs copropagating along the $-z$ direction ($\theta =180^{\circ}$) for the three previous pump wavelengths, as a function of the pump polarization angle. 
\textbf{f-h} Evolution of the fidelity between the biphoton state propagating along the $\theta =180^{\circ}$ direction and specific target states, when the pump polarization is tuned between $0^{\circ}$ and $90^{\circ}$, and pumping the structure at (\textbf{f}) $\lambda_p=735$ nm, (\textbf{g}) $\lambda_p=780$ nm, and (\textbf{h}) $\lambda_p=830$ nm.
\textbf{i, j} The biphoton emission rate spectra ($d^2N_{\mathrm{pair}}/ dt d\omega_s$, shown as a function of the signal photon frequency/wavelength), as a function of the pump frequency/wavelength (pump polarization is fixed to $y$-polarization and pump intensity is $1\times 10^{9}~\mathrm{W/m^2}$). Here, photon pairs within a numerical aperture $NA=0.8$ are considered, either around the forward $-z$ direction (\textbf{i}) or the backward $+z$ direction (\textbf{j}).}
\label{fig:Fig_6}
\end{figure*}

The fact that in certain detection configurations the generation of polarization-entanglement is independent of the modal content of the nanoresonator offers new opportunities for tailoring the remaining spatial and spectral properties of the photon-pair state, without affecting the polarization-entanglement.  
To demonstrate such a control, we consider the case of a nanocylinder made from an AlGaAs crystal with ${<}011{>}$ orientation (see Fig.~\ref{fig:Fig_6}(a)).
The nanocylinder is of the same dimension as before and is excited by a plane wave, linearly polarized at an angle $\phi_c$ with respect to the $x=x_c$ axis (see Fig.~\ref{fig:Fig_6}(a)).

For $\phi_c=90^\circ$, i.e. pump polarization along $y$, the emission diagrams corresponding to copropagating, non-degenerate pairs ($\lambda_s=2\lambda_p-\SI{20}{nm}$) are represented in Fig.~\ref{fig:Fig_6}(b)-(d) for the three pump wavelengths 735, 780, and $\SI{830}{nm}$. Similar to the result obtained for the point-like source under the same pump polarization and detection configuration in Fig.~\ref{fig:Fig_2}(e), a protected $\ket{HV}+\ket{VH}$ polarization-state is generated along the $\pm z$ direction and in the whole $xz$- and $zy$- planes, regardless of the pump wavelength.
Importantly, while the point-like source in Fig.~\ref{fig:Fig_2}(e) always symmetrically radiates the photon pairs into the upper and lower half-spaces, a directional degree of freedom is now added by the highly multi-modal nature of the nanoresonator. Indeed, focusing on the generation diagrams Fig.~\ref{fig:Fig_6}(b-d) obtained for $\phi_c=90^\circ$ we see how most of the emission is redirected from the forward direction ($-z$ axis, Fig.~\ref{fig:Fig_6}(b)) to the backward direction ($z$ axis, Fig.~\ref{fig:Fig_6}(d)) when tuning the pump wavelength. The biphoton polarization state is not affected and remains $\ket{HV} + \ket{VH}$. This directional emission of the photon pairs can be seen as the quantum equivalent of the classical Kerker effect,\cite{nikolaeva2021directional} where complex spatial interferences between the multiple idler and signal QNMs cancel out their co-detection in the forward or backward directions.

Fig.~\ref{fig:Fig_6}(e)-(h) further reveals that in the ${<}011{>}$ crystal, the pump polarization can be used as a degree of freedom to switch from this protected Bell state for pump excitation along the $y$-axis to different entangled or un-entangled polarization states for pump excitation along the $x$-axis. The detection direction is fixed to $\theta=180^\circ$ in these four plots. A similar tuning of the entanglement degree using the pump polarization has also been experimentally observed for an ultra thin Zinc-Blende-type crystal.\cite{sultanov2022flat} The corresponding Schmidt numbers obtained when tuning the pump polarization are reported in Fig.~\ref{fig:Fig_6}(e) for different pump wavelengths. Note that the entangled state $\ket{HV}+\ket{VH}$ obtained for $y$-polarized excitation is independent of the pump wavelength, since it is mediated by the cross-polarized tensor components $\chi^{(2)}_{xyy}$ and $\chi^{(2)}_{yxy}$, which always excite pairs of orthogonally polarized modes. Opposed to that, for $x$-polarized excitation, the tensor elements $\chi^{(2)}_{yyx}$ and $\chi^{(2)}_{zzx}$ couple co-polarized modes. Here, the collective interference of all modes governs the resulting polarization state, which can be controlled by the pump wavelength. The switching in the emission directionality is in general a broad-band effect, as supported by the signal photon spectra presented in Fig. \ref{fig:Fig_6}(i,j). The spectra are calculated for varying pump wavelengths at $\phi_c=90^\circ$ and for copropagating signal and idler photons detected within a 0.8 numerical aperture around the forward (Fig.~\ref{fig:Fig_6}(i)) and backward (Fig.~\ref{fig:Fig_6}(j)) directions.

\section{Discussion and Conclusion}\label{sec12}

In summary, we have shown through both, analytical calculations and numerical simulations, that nanoscale nonlinear systems like point-nanoparticles and cylindrical nanoresonators can naturally generate various maximally entangled polarization (Bell) states. As a specific example, we have chosen materials with Zinc-Blende structure. We uncovered a regime, in which a fixed polarization-entangled Bell state is emitted in all emission directions. This is enabled by the Zinc-Blende type nonlinear tensor and a particular detection configuration. We derived an analytical explanation for this effect in the limit of a point-like nonlinear source and further evidenced that the same behavior is obtained for a finite size nanoresonator. This in principle creates a highly spatially multimode biphoton state that is maximally entangled in the polarization degree of freedom and is also naturally frequency entangled in a very broad spectral range. Our results show that nonlinear nanoresonators are a natural candidate for generation of multimode hyperentangled biphoton states, \cite{sultanov2022flat,barreiro2005} which could find applications in free-space quantum communication protocols, e.g. for superdense coding. 
We also identified regimes in which the Bell state generation in the nanoresonator is independent of the pumping wavelength and/or polarization. This protection allows to engineer the emission directionality and spectral properties of a biphoton state, without affecting its maximally-entangled polarization state.
We showed that in other configurations, where the polarization-entangled state is not protected, properties such as the polarization or wavelength of the pump beam can be used to change the modal content of the nanoresonator. This allows to generate a wide variety of entangled or fully separable biphoton polarization states, with different directional patterns and spectral properties. Overall, our results represent a comprehensive analysis of the physics of photon-pair generation in nanoscale systems. From the application side, our analysis shows that nonlinear nanoresonators are efficient and highly versatile sources of photon pairs with unique properties in generation of complex entangled biphoton states.
On the fundamental side, we uncover a seemingly "protected" regime of entanglement generation.
Our analysis shows that the presence of this effect depends, at least, on the combination of the following factors: type of the detection configuration, the nonlinear properties of the system (such as the orientation of the nonlinear susceptibility tensor), and the linear properties of the system (such as the GF of the system at the signal and idler frequencies being equal or not).
At this stage, we believe that our results suggest the possible existence of a regime of "symmetry-protected entanglement generation", where the symmetry of the system is determined by the combination of these linear, nonlinear, and detection configuration properties. More analysis is to be done in this regard. 
Further analysis could potentially be carried out based on group theory, which has been applied for a symmetry-based description of classical nonlinear parametric processes in nonlinear nanoresonators.\cite{frizyuk2019a,frizyuk2019b} Yet such a formalism should be updated to include the effect of quantum interference, which as we demonstrated, plays the key role in generation of the polarization-entangled states. 

Finally, we point out the recent discovery of the concept of symmetry-protected interaction/scattering of entangled photonic states,\cite{buse2018,lasa2020} which could have potential use for creating decoherence-free subspaces, and we believe it shares fundamental physical connections to what we have demonstrated for generation of entangled states. In fact, a potential combination of the concepts of symmetry-protected generation of entangled states of light and symmetry-protected interaction/scattering of entangled states can pave the way towards creating quantum photonic systems that are highly resilient to decoherence effects.

\section*{Supplementary Material}

See the supplementary material for details on the implementation of the quantum state tomography (section S1) and an example derivation for the photon-pair generation rate (section S2) and condition for polarization-entanglement (section S3) of a point-like nonlinear source. In section S4, a convergence test for the QNM reconstruction of the generated quantum state is shown. Section S5 provides details on the total pair generation rate for a <011> AlGaAs nanoresonator whereas section S6 demonstrates the influence of varying pump-wavelength and pump polarization for a <100> AlGaAs nanoresonator.

\section*{Acknowledgements}

The authors acknowledge the following fundings: Deutsche Forschungsgemeinschaft (DFG, German Research Foundation) within the international research training group GRK2675 and under the project identifiers PE 1524/13-1 (NanoPair), 398816777-SFB 1375 (NOA); Agence Nationale de la Recherche (ANR, French Research Foundation) under the project identifier ANR-18-CE92-0043 (NanoPair); European Union by project METAFAST-899673-FETOPEN-H2020; German Federal Ministry of Education and Research (BMBF) under the project identifiers 13N14877 (QuantIm4Life), 13N16108 (PhoQuant); Thuringian Ministry for Economy, Science, and Digital Society and the European Social Funds (2021 FGI 0043 -Quantum Hub Thuringia). S.S. acknowledges funding by the LIGHT profile (FSU Jena).

\section*{Author Declarations}

\subsection*{Conflict of Interest}

The authors have no conflicts to disclose.

\subsection*{Author Contributions}

Maximilian A. Weissflog and Romain Dezert contributed equally to this work. Adrien Borne and Sina Saravi jointly supervised this work.

\textbf{Maximilian~A.~Weissflog}: Data Curation (equal), Formal Analysis (equal), Methodology (equal), Software (equal), Validation (equal), Visualization (equal), Writing/Original Draft Preparation (equal), Writing/Review \& Editing (equal). \textbf{Romain~Dezert}: Data Curation (equal), Formal Analysis (equal), Methodology (equal), Software (equal), Validation (equal), Visualization (equal), Writing/Original Draft Preparation (equal), Writing/Review \& Editing (equal). \textbf{Vincent~Vinel}: Conceptualization (support), Methodology (support), Writing/Review \& Editing (support). \textbf{Carlo~Gigli}: Data curation (support), Conceptualization (support), Methodology (support), Software (support). \textbf{Giuseppe~Leo}: Conceptualization (equal), Funding Acquisition (equal), Project Administration (support), Resources (equal), Writing/Review \& Editing (support). \textbf{Thomas~Pertsch}: Conceptualization (equal), Funding Acquisition (equal), Project Administration (support), Resources (equal), Writing/Review \& Editing (support). \textbf{Frank~Setzpfandt}: Conceptualization (equal), Funding Acquisition (equal), Project Administration (support), Resources (support), Validation (support), Writing/Review \& Editing (equal). \textbf{Adrien~Borne}: Conceptualization (equal), Methodology (support), Supervision (equal), Validation (equal), Writing/Original Draft Preparation (equal), Writing/Review \& Editing (equal). \textbf{Sina~Saravi}: Conceptualization (equal), Methodology (support), Supervision (equal), Validation (equal), Writing/Original Draft Preparation (equal), Writing/Review \& Editing (equal)

\section*{Data Availability Statement}
The data that support the findings of this study are available from the corresponding author upon reasonable request.

\appendix

\section{Green's Function Method for Description of Photon-Pair Generation}\label{sec:GF_method_photon_pairs}

 With a nonlinear structure pumped with a single-frequency pump beam $\vec{E}_{p} (\vec{r},t)= \vec{E}_{p} (\vec{r}) \exp{-i \omega_p t} + \mathrm{c.c.}$, the rate of photon pairs that can be detected in the far-field follows the relation:\cite{poddubnyGenerationPhotonPlasmonQuantum2016}
\begin{equation}
    \begin{aligned}
         \frac{d^4N_\mathrm{pair}}{dt d\Omega_i d\Omega_s d\omega_s}
        =&\frac{8}{\pi} n_i n_s r_i^2 r_s^2
        \frac{(\omega_p-\omega_s)^3\omega_s^3}{c^6} \\
        &\times\lvert\widetilde{T}_{is} (\vec{r}_i,\omega_p-\omega_s,\vec{e}_i;\vec{r}_s,\omega_s,\vec{e}_s) \rvert^2
        \label{N_solid_without2pi}
    \end{aligned}
\end{equation}

with the two photon amplitude
\begin{equation}
    \begin{aligned}
         \widetilde{T}_{is}
         =&
         \sum_{\alpha,\beta,\gamma,q_i,q_s}
    e_{i,q_i} e_{s,q_s} \\
    &\times\int d\vec{r} \:
    \chi^{(2)}_{\alpha\beta\gamma} (\vec{r}) \:
    E_{p,\gamma} (\vec{r})
    G_{q_i\alpha}(\vec{r}_i,\vec{r},\omega_i)
    G_{q_s\beta}(\vec{r}_s,\vec{r},\omega_s). 
    \label{T_is_tilde}
    \end{aligned}
\end{equation}
This relation can be found in the supplementary of Ref.\cite{poddubnyGenerationPhotonPlasmonQuantum2016}, where we updated the pre-factors from the Gaussian units to the SI units, matching our definition of the GF, and also starting with the nonlinear Hamiltonian $\hat{H}_\mathrm{NL} (t) = - \varepsilon_0 \int d^3 \vec{r} \sum_{\alpha,\beta,\gamma} \chi^{(2)}_{\alpha \beta \gamma} (\vec{r}) \hat{E}^{(-)}_{\alpha}(\vec{r}) \hat{E}^{(-)}_{\beta}(\vec{r})
E_{p,\gamma}(\vec{r})\allowbreak \exp{-i\omega_p t} + {\rm H.c.}$ (H.c. is Hermitian conjugate). The detection rate is given per units of far-field solid angles $\Omega_i$ (for the idler photon) and $\Omega_s$ (for the signal photon), and per unit of signal-photon angular frequency $\omega_s$. The idler angular frequency is fixed by conservation of energy to $\omega_p-\omega_s$. The detection polarization for the idler and signal photons are along the $\vec{e}_{i}$ and $\vec{e}_{s}$ directions, respectively. The indices run over the Cartesian coordinates, $\left(\alpha,\beta,\gamma,q_i,q_s\right)=\{x,y,z\}$. $\vec{r}_i$ and $\vec{r}_s$ are the positions for the detection of the idler and signal photons in the far-field, respectively, with $n_i$ and $n_s$ being the refractive indices of the detection medium in the far-field at the idler and signal frequencies. Here, we consider it to be free space $n_i=n_s=1$. The GF tensor $\bar{G}$ is the solution to the Helmholtz equation
$\left[ \nabla\times\nabla\times- \frac{\omega^2}{c^2} \varepsilon (\vec{r},\omega) \right] \bar{G} (\vec{r},\vec{r}',\omega) = \bar{I} \delta (\vec{r}-\vec{r}')$. For the scenario with a point-like nonlinear source, we use the free-space GF in Eq.~\eqref{T_is_tilde}, which has an analytical form\cite{Nov06} (also see supplementary section S2 for an example derivation).

\section{Polarization Quantum-State Retrieval using Tomographic Method}\label{sec:Q_state_tomography}

The GF quantization method makes use of local bosonic excitation operators,\cite{Kno00} which do not directly translate to photon operators associated to optical normal modes. Hence, the method for description of photon-pair-generation based on the GF quantization method in\cite{poddubnyGenerationPhotonPlasmonQuantum2016} does not predict the quantum state of the photon pair in certain normal modes, but rather predicts the probabilities for detecting photon pairs at certain locations with certain frequencies and polarizations.
To extract the polarization quantum state, we use a polarization tomography method, which is commonly used in experimental situations where one measures probabilities for detecting photon pairs with certain properties. We explain this method shortly in the following, and in more detail in supplementary section S1.

Density matrices of the biphoton polarization states are determined following the tomographic procedure described in Ref.\cite{james2001measurement}. The density matrix $\hat{\rho}$ expressed in the $\{\ket{HH},\ket{HV},\ket{VH},\ket{VV}\}$ basis is retrieved from projective measurements computed  for a set of 16 tomographic states $\{\ket{\psi_\nu}\}_{\nu \in [\![1;16]\!]} $. We use the same set of probe states as proposed in Ref.\cite{james2001measurement} (see supplementary section S1).
The linear tomographic reconstruction can be written under this compact form:
\begin{equation}\label{eq:finalexpansion}
   \hat{\rho}= \dfrac{1}{\mathcal{N}} \sum_{\nu=1}^{16} n_\nu  \hat{M}_{\nu}
\end{equation}
where $(\hat{M}_{\nu})_\nu$ is a basis of $4\times4$ matrices set by the choice of probe states, $\mathcal{N}$ is a normalisation constant and $n_\nu$ are the weights to be evaluated for the reconstruction. These coefficients correspond to the coincidence rate obtained for each probe state:
\begin{equation}\label{eq:tomocoef}
n_\nu=\dfrac{d^4N_{\mathrm{pair}}}{dtd\omega_sd\Omega_id\Omega_s}(\theta_i,\varphi_i,\omega_i,\vec{d}^\nu_i;\theta_s,\varphi_s,\omega_s,\vec{d}^\nu_s)
\end{equation}
where the detected polarization directions $(\vec{d}^\nu_s,\vec{d}^\nu_i)$ are aligned according to the state $\nu$ to probe.
We access the polarization state of the pairs for any detection configuration by repeating the tomographic reconstruction procedure for each possible emission direction of the signal and idler photons. The retrieved polarization states corresponding to a given set of signal and idler directions are pure states (verifying $Tr(\hat{\rho})=1,\forall \left( (\varphi_s,\theta_s),(\varphi_i,\theta_i) \right) $).

The final form of the retrieved density matrices depends on the convention adopted to define the basis vectors $\ket{H}$ and $\ket{V}$. 
The entire polarization analysis carried out in this work is conducted by defining $\ket{H}=\Vec{e'}_x$ and $\ket{V}=\Vec{e'}_y$ where 
\begin{equation}
     \begin{cases}
    \Vec{e'}_x&=\cos{\varphi}\, \Vec{e}_\theta-\sin{\varphi}\, \Vec{e}_\varphi \\
    \Vec{e'}_y&=\sin{\varphi}\, \Vec{e}_\theta+\cos{\varphi}\, \Vec{e}_\varphi
         \end{cases}
\end{equation}
in the upper half space ($z>0$) and 
\begin{equation}
     \begin{cases}
    \Vec{e'}_x&=-\cos{\varphi}\, \Vec{e}_\theta-\sin{\varphi} \, \Vec{e}_\varphi\\
    \Vec{e'}_y&=-\sin{\varphi} \, \Vec{e}_\theta+\cos{\varphi} \,\Vec{e}_\varphi
     \end{cases}
\end{equation}
in the lower half space ($z<0$).
These vectors are defined so that they would coincide with the $\Vec{e}_x$ and $\Vec{e}_y$ vectors of the laboratory reference frame if we were to apply a rotation corresponding to a fictitious collection and collimation scheme of the pairs generated in all directions of space via two lenses of numerical aperture 1, placed above ($+z$ half-space) and below ($-z$ half-space) the nonlinear structure (see supplementary section S1 for detailed information). This basis choice therefore provides an analysis and interpretation of the polarization state in correspondence with a potential experimental measurement.

Throughout the main text, the fidelities between two states (characterized by density matrices $\hat{\rho}$ and $\hat{\sigma}$) are calculated as:\cite{nielsen_chuang_2010,UHLMANN1976273}
\begin{equation}
F(\hat{\rho},\hat{\sigma})=\left[Tr\left(\sqrt{\sqrt{\hat{\rho}}\,\hat{\sigma}\sqrt{\hat{\rho}}}\right)\right]^2
 \end{equation}
The Schmidt entanglement parameter $K$ (Schmidt number) quantifying the degree of entanglement of the pair is given by:\cite{grobe1994measure,eberly2006schmidt}
\begin{equation}
    K= \dfrac{1}{Tr(\hat{\rho}_i^2)}=\dfrac{1}{Tr(\hat{\rho}_s^2)}
\end{equation}
where $\hat{\rho}_s$ or $\hat{\rho}_i$ are the reduced density matrices corresponding to the single particle wavefunctions which are obtained from the partial trace of the biphoton density matrix: $\hat{\rho}_{s(i)}=Tr_{i(s)}(\hat{\rho})$.

It should be emphasized that the specific shape of the entangled state depends on the definition of the local polarization basis H and V for each of the signal and idler channels, which is an arbitrary choice. 
However, what is independent of the basis choice is the degree of entanglement, given by the Schmidt number. Hence, a Schmidt number of 2 means a maximally entangled state in any polarization basis.

\section{Green's Function Method combined with QNM Expansion for Rigorous Description of Pair-Generation in a Nonlinear Nanoresonator}\label{sec:GF_QNM_method}

We insert an expansion of the GF based on QNMs:\cite{Lal18}
\begin{align}
    G_{\alpha\beta}(\vec{r},\vec{r}',\omega)= \frac{-1}{\mu_0} \sum_{m=1}^{\infty} \frac{ \widetilde{E}_{m,\alpha}(\vec{r}) \times \widetilde{E}_{m,\beta}(\vec{r}') }{ (\omega-\widetilde{\omega}_m)\widetilde{\omega}_m } 
    \label{eq2_1}
\end{align}
into Eq.~\eqref{T_is_tilde}. Here $\widetilde{E}_{m,q}$ is the $q$-polarization of the electric field profile of the $m^\mathrm{th}$ quasinormal mode, where $\widetilde{\omega}_m$ is the complex-valued resonance frequency of that QNM. 
Notice that this expansion is valid together with the following normalization relation:\cite{Lal18}
\begin{equation}
\int d\vec{r} \: \left[ \vec{\widetilde{E}}_m(\vec{r}) \cdot \varepsilon_0 \frac{\partial \omega \varepsilon}{\partial \omega} \vec{\widetilde{E}}_m(\vec{r}) -\vec{
\widetilde{H}}_m(\vec{r}) \cdot \mu_0 \frac{\partial \omega \mu}{\partial \omega} \vec{\widetilde{H}}_m(\vec{r})
\right] = 1   ,
    \label{eq22_1}
\end{equation}
where the derivatives are evaluated at the complex resonance frequencies $\widetilde{\omega}_m$. $\vec{
\widetilde{H}}_m(\vec{r})$ is the magnetic field profile of the $m^\mathrm{th}$ quasinormal mode. $\varepsilon_0$ ($\varepsilon$) and $\mu_0$ ($\mu$) are the vacuum (relative) permittivity and permeability. This integral also includes the volumes of the perfectly matched layers (PMLs).

Note that since in general the QNMs for such open systems do not obey the conventional power-orthogonality conditions of closed optical systems such as a lossless waveguide or a high-Q resonator, the contribution of different QNMs to pair-generation cannot be separated from each other and all cross-QNM combinations have to be considered at once in the SPDC process.
After a straightforward rearrangement of the terms we therefore find:
\begin{equation}
    \begin{aligned}
\widetilde{T}_{is} &(\vec{r}_i,\omega_p-\omega_s,\vec{e}_i;\vec{r}_s,\omega_s,\vec{e}_s) 
        = 
        \frac{1}{\mu_0^2} \sum_{q_i,q_s}
        e_{i,q_i} e_{s,q_s} \\
        &\times\sum_{m,n=1}^{\infty}\xi_{m,n}(\omega_s, \omega_p) \widetilde{E}_{m,q_i}(\vec{r}_i)\widetilde{E}_{n,q_s}(\vec{r}_s),
        \label{eq:SPDC_rate}
    \end{aligned}
\end{equation}
where the contribution of each pair of QNMs $m,n$ to the down-conversion process is captured by the dispersive modal overlap factor $\xi_{m,n}(\omega_s, \omega_p)$:
\begin{equation}
    \begin{aligned}\xi_{m,n}(\omega_s, \omega_p)=&\frac{\sum_{\alpha,\beta,\gamma} \int d\vec{r} \:
    \chi^{(2)}_{\alpha\beta\gamma} (\vec{r}) \widetilde{E}_{m,\alpha}(\vec{r}) \widetilde{E}_{n,\beta}(\vec{r})
    E_{p,\gamma} (\vec{r})}{ (\omega_p-\omega_s-\widetilde{\omega}_m)\widetilde{\omega}_m (\omega_s-\widetilde{\omega}_n) \widetilde{\omega}_n } .
    \label{eq:overlap_xi}
    \end{aligned}
\end{equation}

For our numerical calculations, we obtain the pump field $\vec{E}_{p} (\vec{r})$ directly by rigorous simulations of Maxwell's equations in COMSOL Multiphysics, through exciting the nanoresonator with a plane wave at frequency $\omega_p$ and calculating the resulting electric field in the nanoresonator. This can also be done more analytically, by also expanding the expression for the classical pump field into the QNMs of the system.

\section{Computing QNM Near- and Far-fields}\label{sec:computing_near_far_fields}

In order to find the resonator QNMs, the eigenvalue problem posed by the source-free Maxwell's equations is solved with the commercial finite element method based software COMSOL Multiphysics using an openly available toolbox.\cite{wuModalAnalysisElectromagnetic2023} Since for the general case of a dispersive material the eigenvalue problem is nonlinear, we follow the approach of\cite{yanRigorousModalAnalysis2018,wuModalAnalysisElectromagnetic2023} and linearize the eigenvalue problem using auxiliary fields. The appropriate normalization of the divergent QNM fields is performed by bounding them with PMLs and computing a volume integral over the entire computational domain, including both the PML region and the inner physical domain.\cite{sauvanTheorySpontaneousOptical2013,wuModalAnalysisElectromagnetic2023} 
QNMs of purely numerical origin \cite{yanRigorousModalAnalysis2018}, mainly associated to the PMLs, do practically not contribute to the overlap integral over the nonlinear source Eq.~\eqref{eq:overlap_xi}, which automatically ensures that all the analysis regarding the physics of pair-generation in a nanoresonator rely on the physical resonances of the system.

The photon-pair detection and polarization analyses are carried out in the far-field where the fields evolve as $\vec{E}(\vec{r})=(e^{iknr}/r)\vec{E}(\theta,\varphi)$. We obtain the far-field angular radiation pattern of the QNMs $\widetilde{E}_m(\theta,\varphi)$ by applying a numerical near-to-far-field transformation (NFFT) to the normalized QNM near-fields. This transformation also applies to inhomogeneous backgrounds.\cite{yangNeartoFarFieldTransformations2016} We performed the NFFT for all modes once at a same single wavelength taken in the center of the SPDC spectral range of interest.\cite{kongsuwanPlasmonicNanocavityModes2020} While strictly speaking this transformation would need to be carried out for each mode at each SPDC wavelength investigated, choosing a single transformation wavelength for all QNM is a common approach due to the weak dependence of the modal far-field on the transformation wavelength.\cite{wuModalAnalysisElectromagnetic2023}

\section*{References}

\bibliography{aipsamp}

\end{document}